\documentclass[12pt]{article}
\usepackage[english]{babel}
\usepackage[latin1]{inputenc}
\usepackage{amsfonts,amssymb,amsmath, epsfig}
\usepackage{color,graphicx,graphics,psfrag}
\usepackage{amsmath,amstext,amssymb,amsfonts, amscd}
\usepackage{hyperref}           % faire un pdf avec des liens dynamiques

\def\Box{\leavevmode\vbox{\hrule
     \hbox{\vrule\kern4pt\vbox{\kern4pt}%
           \vrule}\hrule}}
\def\blackbox{\leavevmode\vrule height 5pt width 4pt depth 0pt\relax}
\def\endproof{\null\hfill {$\blackbox$}\bigskip}

\newcounter{appendix}
\def\appendix{\advance\c@appendix by 1
   \def\thesection{\Alph{section}}
   \ifnum\c@appendix=1 \setcounter{section}{-1} \fi
   \@startsection {section}{1}{\z@}{-3.5ex plus -1ex minus 
   -.2ex}{2.3ex plus .2ex}{\Large\bf}}

\def\paragraph#1{{\bf #1\ }}

\newtheorem{lemma}{Lemma}[section]  

\newtheorem{theorem}[lemma]{Theorem}

\newtheorem{remark}{Remark}[section]
\title{Kinetic formulation and global existence for the Hall-Magneto-hydrodynamics system} 
\author{M. Acheritogaray$^{(1,2)}$, P. Degond$^{(1,2)}$, A. Frouvelle$^{(1,2)}$, J-G. Liu$^{(3)}$} 
\date{} 
\begin{document}

\maketitle

\vspace{0.5 cm}

\begin{center}
1-Université de Toulouse; UPS, INSA, UT1, UTM ;\\ 
Institut de Mathématiques de Toulouse ; \\
F-31062 Toulouse, France. \\
2-CNRS; Institut de Mathématiques de Toulouse UMR 5219 ;\\ 
F-31062 Toulouse, France.\\
email: marion.acheritogaray@math.univ-toulouse.fr, pierre.degond@math.univ-toulouse.fr, amic.frouvelle@math.univ-toulouse.fr
\end{center}

\begin{center}
3- Department of Physics and Department of Mathematics\\
Duke University\\
Durham, NC 27708, USA\\
email: jliu@phy.duke.edu
\end{center}

\vspace{0.5 cm}
\begin{abstract}
This paper deals with the derivation and analysis of the the Hall Magneto-Hydrodynamic equations. We first provide a derivation of this system from a two-fluids Euler-Maxwell system for electrons and ions, through a set of scaling limits. We also propose a kinetic formulation for the Hall-MHD equations which contains as fluid closure different variants of the Hall-MHD model. Then, we prove the existence of global weak solutions for the incompressible viscous resistive Hall-MHD model. We use the particular structure of the Hall term which has zero contribution to the energy identity. Finally, we discuss particular solutions in the form of axisymmetric purely swirling magnetic fields and propose some regularization of the Hall equation. 
\end{abstract}

\medskip
\noindent
{\bf Acknowledgements:} This work has been supported by the 'Fondation Sciences et Technologies pour 
l'Aéronautique et l'Espace', in the frame of the project 'Plasmax' 
(contract \# RTRA-STAE/2007/PF/002) who supported the fourth author's stay in Toulouse. This work has also been supported by the 'Commissariat  \`a l'\'Energie Atomique (CEA)' 
in the frame of the contract 'PICCADI' (contract \# 4600192523). 
The second and fourth authors wish to acknowledge the hospitality of Mathematics Department  and Mathematical Sciences Center of Tsinghua University where this research was completed. The research of J.-G. L. was partially supported by NSF grant DMS 10-11738. The second author wishes to thank F. Deluzet, G. Gallice and C. Tessieras for fruitful discussions.


\vskip 0.4cm

%%%%%%%%%%%%%%%%%%%%%%%%%%%%%%%%%%%%%%%%%%%%%%%%%%%%%%%%%%%%%%%%%%%%%%%%%%%%%%%%%%%%%%%%%%%%%%%%
%%%%%%%%%%%%%%%%%%%%%%%%%%%%%%%%%%%%%%%%%%%%%%%%%%%%%%%%%%%%%%%%%%%%%%%%%%%%%%%%%%%%%%%%%%%%%%%%
%%%%%%%%%%%%%%%%%%%%%%%%%%%%%%%%%%%%%%%%%%%%%%%%%%%%%%%%%%%%%%%%%%%%%%%%%%%%%%%%%%%%%%%%%%%%%%%%
%%%%%%%%%%%%%%%%%%%%%%%%%%%%%%%%%%%%%%%%%%%%%%%%%%%%%%%%%%%%%%%%%%%%%%%%%%%%%%%%%%%%%%%%%%%%%%%%
\setcounter{equation}{0}
\section{Introduction}
\label{sec:intro}

This paper deals with the derivation and analysis of the the Hall Magneto-Hydrodynamic (Hall-MHD) equations. The Hall-MHD model currently receives an increasing attention from plasma physicists. It is believed to be the key for understanding the problem of magnetic reconnection. Indeed, space plasma observations provide strong evidence for the existence of frequent and fast changes in the topology of magnetic field lines, associated to violent events such as solar flares \cite{Forbes}. However, magnetic reconnection cannot be described in the framework of ideal MHD, due to the frozen-field effect. Indeed, in ideal MHD, due to the Faraday equation and ideal Ohm's law, the magnetic field is essentially passively transported by the fluid velocity. Therefore, the topology of the magnetic field is preserved, even in the magnetic field lines are deformed by the flow. In order to break this passive magnetic field transport by the fluid flow, one is led to re-introduce the Hall terms which was neglected in ideal MHD. In spite of its increasing importance for physical applications, the Hall-MHD model has received very little attention from the theoretical viewpoint (see e.g. \cite{MK, SC}) and the purpose of this paper is mainly to propose a framework for the derivation and analysis of the Hall-MHD problem. 

We first provide a derivation of this system from a two-fluids isothermal Euler-Maxwell system for electrons and ions, through a set of scaling limits. The two-fluids model of plasma is known as Braginskii model \cite{Braginskii} and was justified in \cite{DL} (see also \cite{Deg_plasmas}) on the basis of a concurrent hydrodynamic and zero electron to ion mass ratio limit. As usual in MHD models, zero electron to ion mass ratio limit, zero Debye length limit and zero displacement current limits have to be taken. Then, the main point is to examine the orders of magnitude of the various terms arising in the generalized Ohm law (which is the electron momentum equation with zero inertia) and in the current equation. Basically, Hall MHD is obtained when the electron and ion velocities have difference of order unity and when this difference is introduced inside the generalized Ohm law.  

Then, we propose a kinetic formulation for the Hall-MHD equations which contains as fluid closure different variants of the Hall-MHD model. The kinetic formulation consists of a Fokker-Planck equation for the ions and a set of fluid equations for the electrons coupled through quasineutrality. The Fokker-Planck operator models electron-ion collisions and contributes to relaxation of the velocities and the temperatures of both species to a common value. This kinetic model was in particular justified in \cite{Deg_Luc}. The MHD equations are obtained by taking the fluid moments of the ion Fokker-Planck equation and closing the resulting equations by a Maxwellian assumption. The resulting two-temperature resistive Hall MHD model consists of conservation equations for the density, momentum, energy and magnetic field combined with an evolution equation for the electron temperature and with the generalized Ohm law. The proposed hybrid ion-kinetic, electron-fluid model bear strong analogies with models used in the literature for numerical simulations such as \cite{VTCHM}, but the purpose is to highlight its mathematical structures. 
Indeed, a particularly interesting special case is when the electron and ion temperatures are equal to the same constant value (isothermal single temperature resistive Hall MHD). In this case, we can rephrase its kinetic formulation in the form of a coupled Fokker-Planck Faraday system, which exhibits an entropy dissipation identity. Surprisingly enough, the kinetic formulation of standard ideal MHD is deduced by neglecting the Hall term in Faraday's equation but keeping it in the kinetic equation. 

The theoretical analysis focuses on  the existence of global weak solutions for the incompressible viscous resistive Hall-MHD model written as follows:
\begin{eqnarray}
& & \hspace{-1cm}  \partial_t u +  u \cdot \nabla u + \nabla p  =  (\nabla \times B) \times B + \Delta u,  \label{eq:dtu} \\
& & \hspace{-1cm} \nabla \cdot u = 0 , \label{eq:divu}\\
& & \hspace{-1cm} \partial_t B -  \nabla \times (u \times B) + \nabla \times ( (\nabla \times B) \times B)  = \Delta B , \label{eq:dtB}\\
& & \hspace{-1cm} \nabla \cdot B = 0 , \label{eq:divB}
\end{eqnarray}
where $u(x,t)$ and $B(x,t)$ are the fluid velocity and magnetic field, depending on the spatial position $x$ and the time $t$. The result is valid on a square domain $\Omega$ of ${\mathbb R}^3$ with periodic boundary conditions. We stress that the important contribution of this work is the account of the last term of the left-hand side of (\ref{eq:dtB}), known as the Hall effect term. The main theorem of this work is stated as follows: 

\begin{theorem}
Let $\Omega = [0,1]^3$. Assume that $u_0 \in (L^2(\Omega ))^3$,  $B_0 \in (L^2(\Omega ))^3$ with $\nabla \cdot u_0=0$, $\nabla \cdot B_0=0$. Then, there exists a global weak solution $(u,B)$ for the Hall MHD problem (\ref{eq:dtu}), (\ref{eq:divB}). Moreover, we have $(u,B) \in L^{\infty }((0,T),L^2(\Omega))$ $ \cap $ $ L^2((0,T),$ $H^1(\Omega))$ and $\partial_t u \in L^{\frac{4}{3}}((0,T),H^{-1}(\Omega))$, $\partial_t B \in L^{\frac{4}{3}}((0,T),H^{-2}(\Omega))$. Additionally, the following energy inequality holds:
\begin{equation}
\frac{d}{dt} {\mathcal E}(t) + \| \nabla u \|^2_{L^2(\Omega)} + \| \nabla B \|^2_{L^2(\Omega)} \leq 0, 
\label{eq:energy_inequality}
\end{equation}
with 
\begin{equation}
{\mathcal E}(t) = \frac{1}{2} \|  u \|^2_{L^2(\Omega)} + \| B \|^2_{L^2(\Omega)} . 
\label{eq:energy}
\end{equation}
\label{th:global}
\end{theorem}

In fact, the main difficulty of this work is concentrated in the treatment of the Hall term. So, we will show a preliminary result for the following Hall problem 
\begin{eqnarray}
& & \hspace{-1cm} \partial_t B + \nabla \times ( (\nabla \times B) \times B)  = \Delta B , \label{eq:dtB2}\\
& & \hspace{-1cm} \nabla \cdot B = 0 , \label{eq:divB2}
\end{eqnarray}
and will provide a detailed proof. The proof uses the particular structure of the Hall term which has zero contribution to the energy identity. The proof of the existence for the coupled system is then a direct consequence of the energy inequality (\ref{eq:energy_inequality}) and will only be sketched. This result is up to our knowledge the first theoretical result for Hall MHD. Global existence for standard viscous resistive incompressible MHD has been previously proved by Duvaut \& Lions \cite{DL}. A bifurcation analysis of the Hall-MHD problem in view of the question of magnetic reconnection is performed in \cite{HG}. Numerical methods for solving the Hall-MHD problems can be found e.g. in \cite{ADG, CK, HM, HR, KLBD}

\begin{remark}
In a general domain $\Omega$, the physically relevant boundary condition is  the perfectly conducting wall boundary condition, which consists in assuming zero normal component of the $B$ field and zero tangential component of the electric field. Here, the electric field is the quantity inside the curl operator, namely 
$$ E = (\nabla \times B)\times B + \nabla \times B. $$
This leads to the nonlinear boundary conditions
\begin{equation*}
\left\{ 
\begin{aligned}
&B\cdot n=0&\quad & \text{ on } \partial \Omega, \\
&n\times (\nabla \times B) + n \times ( (\nabla \times B)\times B  ) =0&\quad & \text{ on }\partial \Omega . \\
\end{aligned}
\right.
\end{equation*} 
From these conditions, we deduce that 
$$ B \cdot (\nabla \times B) = 0, $$
which means that there is no helicity on the boundary. Because of the nonlinearity of this boundary condition, the methods developed below do not apply. 
\end{remark}

Finally, we discuss particular solutions of the Hall problem in the form of axisymmetric purely swirling magnetic fields and propose some regularization of the Hall equation. For axisymmetric purely swirling magnetic fields, the Hall problem reduces to a viscous Burger's equation. By neglecting the resistivity, the resulting inviscid Burger's equation shows shock wave solutions which are known in physics textbooks as KMC waves for Kingsep, Mokhov and Chukbar \cite{KMC} (see also \cite{CG, DRG}). They only exist if the Hall term is present. This effect also generates boundary layers which lead to nonlinear boundary conditions (see e.g. \cite{MR}). Focusing on the non-resistive Hall problem itself, we propose a regularization consisting in restoring the displacement current in the Ampere equation. We then provide two equivalent formulations of this regularized problem which are obtained when either the current or the electric fields are eliminated from the system. 

The organization of the paper is as follows. In Section \ref{sec:derivation}, we propose a derivation of the Hall-MHD model from the two-fluids Euler-Maxwell model under suitable scaling hypotheses. Then, Section \ref{sec:derivation_kinetic} is devoted to the presentation of the kinetic formulation of the Hall-MHD problem. Section \ref{sec:existence} is focuses on the proof of the existence of global weak solutions for the incompressible viscous resistive Hall-MHD equations. Section \ref{sec:axi} discusses the particular case of axisymmetric purely swirling magnetic fields and proposes a regularization of the Hall problem by means of a re-introduction of the displacement current in the Ampere equation. Finally, a conclusion is drawn in Section \ref{sec:conclu}.

%%%%%%%%%%%%%%%%%%%%%%%%%%%%%%%%%%%%%%%%%%%%%%%%%%%%%%%%%%%%%%%%%%%%%%%%%%%%%%%%%%%%%%%%%%%%%%%%
%%%%%%%%%%%%%%%%%%%%%%%%%%%%%%%%%%%%%%%%%%%%%%%%%%%%%%%%%%%%%%%%%%%%%%%%%%%%%%%%%%%%%%%%%%%%%%%%
%%%%%%%%%%%%%%%%%%%%%%%%%%%%%%%%%%%%%%%%%%%%%%%%%%%%%%%%%%%%%%%%%%%%%%%%%%%%%%%%%%%%%%%%%%%%%%%%
%%%%%%%%%%%%%%%%%%%%%%%%%%%%%%%%%%%%%%%%%%%%%%%%%%%%%%%%%%%%%%%%%%%%%%%%%%%%%%%%%%%%%%%%%%%%%%%%
\setcounter{equation}{0}
\section{Derivation of the Hall-MHD equations from fluid equations}
\label{sec:derivation}

In this section, we briefly motivate the derivation of the model we are considering. For simplicity, we consider the compressible inviscid model and later on change to viscous incompressible flow. We start from the two-fluid isothermal Euler-Maxwell system for the electrons and ions, where we assume that the electron and ion temperatures are the same given constant: 
\begin{eqnarray}
& & \hspace{-1cm} \partial_t n _e +   \nabla \cdot (n _e u _e) = 0 , \nonumber \\
& & \\
& & \hspace{-1cm} m_e ( \partial_t (n _e u _e)  +    \nabla ( n _e  u _e \otimes u _e )) + \nabla (n_e T)  = \nonumber  \\
& & \hspace{2cm} = - e n_e  (E + u_e \times B)  - e^2 \eta n_e n_i (u_e - u_i)  , \label{eq:electron_momentum} \\
& & \hspace{-1cm} \partial_t n _i +   \nabla \cdot (n _i u _i) = 0  , \nonumber \\
& & \hspace{-1cm} m_i ( \partial_t (n _i u _i)  +    \nabla ( n _i  u _i \otimes u _i )) + \nabla (n_i T)  = \nonumber  \\
& & \hspace{2cm} = e n_i  (E + u_i \times B)  -  e^2 \eta n_i n_e (u_i - u_e)  , \nonumber \\
& & \hspace{-1cm}	c^{-2} \partial_t E -  \nabla \times B = - \mu_0 j , \nonumber \\
& & \hspace{-1cm}  \epsilon_0 \nabla \cdot E = \rho , \nonumber \\
& & \hspace{-1cm} \partial_t B +  \nabla \times E = 0 \, , \nonumber \\
& & \hspace{-1cm}  \nabla \cdot B = 0 , \nonumber \\
& & \hspace{-1cm} \rho = e (n_i - n_e) , \nonumber \\
& & \hspace{-1cm} j = e (n_i u_i - n_e u_e)  . \nonumber 
\end{eqnarray}
where $n_e$ and $n_i$ are the electron and ion densities, $u_e$ and $u_i$, the velocities, $T$, the common electron and ion temperature, $m_e$ and $m_i$ the masses. $e$ denotes the elementary positive charge, and we assume singly charged positive ions. $\eta$ is the resistivity due to the electron-ion collisions. $E$, $B$, $\rho$, $j$ are respectively the electric field, the magnetic field, the charge density and the current density. $\epsilon_0$, $\mu_0$ and $c$ are respectively the vacuum permittivity, the vacuum permeability and the speed of light, related by the relation $\epsilon_0 \mu_0 c^2 = 1$. For simplicity, we assume monoatomic perfect gas equations of states for both the electrons and ions. We make the Boltzmann constant equal to unity which means that we measure temperatures in units of energy. The last terms at the right-hand sides of the second and fourth equations are the contributions of the electron-ion collisions to the momentum equation of each species. The two terms sum up to zero which expresses the conservation of total momentum in such collisions. 

We introduce scaling units $n_0$, $u_0$, $E_0$, $B_0$, $x_0$, $t_0$, $\rho_0$, $j_0$ for respectively the densities, velocities, electric field, magnetic field, space, time, charge and current. We assume that these units are related by the following relations:
$$ x_0 = u_0 t_0, \quad u_0 = \sqrt{\frac{T}{m_i}}, \quad E_0 = u_0 B_0, \quad \rho_0 = e n_0.   $$
The first relation means that we observe the system at the convection time scale. The second relation states that the convection velocity is that of the ion thermal speed. The third relation is typical of a MHD scaling and states that the main contribution to the electric field is induction due to the motion of the charged fluid. Finally, the last relation expresses the consistency between the density and charge units. 

Then, six  dimensionless parameters appear:  
\begin{eqnarray*}
&  & \varepsilon^2 = \frac{m_e}{m_i} \, , \quad \alpha^2 = \frac{e E_0 x_0}{T}\, , \quad  \beta = \frac{e^2 \eta n_0 u_0 x_0}{T}  \\
& & \gamma = \frac{u_0}{c}   \, , \quad \lambda^2 = \frac{\varepsilon_0 T}{e^2 n_0 x_0^2} \, , \quad \eta = \frac{j_0}{e n_0 u_0}.  
\end{eqnarray*}
which have the following interpretation. $\varepsilon^2$ is the electron to ion mass ratio and is very small. $\alpha^2$ is the ratio of the electric energy to the thermal energy. $\beta$ measures the relaxation frequency of the electron and ion velocities due to collisions. $\gamma$ is the ratio of the fluid velocity to the speed of light. $\lambda$ is the scaled Debye length and measures the closedness to quasi-neutrality. $\eta$ is the ratio of the charge current scale to the electron or ion current scales. Since the charge current is the difference of these two particle currents, it may be much smaller than any of them due to charge neutrality. Therefore, the scale ratio $\eta$ may be either $O(1)$ or $\ll 1$ according to the situations.

The dimensionless two-fluids Euler-Maxwell system is written: 
\begin{eqnarray*}
& & \hspace{-1cm} \partial_t n _e +   \nabla \cdot (n _e u _e) = 0 , \\
& & \hspace{-1cm} \varepsilon^2 ( \partial_t (n _e u _e)  +    \nabla ( n _e  u _e \otimes u _e )) + \nabla (n_e T)  = \\
& & \hspace{2cm} = - \alpha^2 n_e  (E + u_e \times B)  - \beta n_e n_i (u_e - u_i)  , \\
& & \hspace{-1cm} \partial_t n _i +   \nabla \cdot (n _i u _i) = 0  , \\
& & \hspace{-1cm}  \partial_t (n _i u _i)  +    \nabla ( n _i  u _i \otimes u _i ) + \nabla (n_i T)  = \\
& & \hspace{2cm} = \alpha^2 n_i  (E + u_i \times B)  - \beta n_i n_e (u_i - u_e) ,  \\
& & \hspace{-1cm}	\gamma^2 \partial_t E -  \nabla \times B = - \frac{\gamma^2\eta}{\alpha^2 \lambda^2 } j , \\
& & \hspace{-1cm}  \alpha^2 \lambda^2 \nabla \cdot E = \rho ,\\
& & \hspace{-1cm} \partial_t B +  \nabla \times E = 0 \, , \\
& & \hspace{-1cm}  \nabla \cdot B = 0 , \\
& & \hspace{-1cm} \rho = n_i - n_e , \\
& & \hspace{-1cm} j = \frac1{\eta} (n_i u_i - n_e u_e)  . 
\end{eqnarray*}

The compressible MHD equation corresponds to the simultaneous independent four limits
\begin{enumerate}
\item $\varepsilon^2 \to 0$: this corresponds to the neglect of the convection term in the electron momentum equation. The resulting equation is usually referred to as the generalized Ohm's law.  
\item $\lambda^2 \to 0$. This gives rise to quasineutrality, i.e. the fact that the local electron and ion densities are everywhere the same. We now denote by $n$ their common value: $n_e = n_i = n$. 
\item $\gamma^2 \to 0$ while keeping $\frac{\gamma^2 \eta}{\alpha^2 \lambda^2 } = 1$. This leads to the neglect of the displacement current in Ampere's equation and gives rise to the standard magnetostatic Ampere law.  
\end{enumerate}

The resulting system is the so-called compressible isothermal resistive Hall-MHD equations. Denoting by $u$ the ion velocity, this system is written
\begin{eqnarray}
& & \hspace{-1cm} \partial_t n +   \nabla \cdot (n u ) = 0 , \nonumber \\
& & \hspace{-1cm}  \partial_t (n  u )  +    \nabla ( n   u  \otimes u  ) + \nabla (2 n T)  =  \alpha^2 \, \eta  \, j \times B ,  \label{eq:momentum} \\
& & \hspace{-1cm} \nabla \times B = j , \nonumber \\
& & \hspace{-1cm} \partial_t B +  \nabla \times E = 0 , \nonumber \\
& & \hspace{-1cm} \nabla \cdot B = 0 , \nonumber \\
& & \hspace{-1cm} j = \frac{1}{\eta} n (u - u_e), \label{eq:current} \\
& & \hspace{-1cm} E + u \times B =  - \frac{T}{\alpha^2}  \nabla ( \mbox{ln} \, n ) + \eta \frac{j \times B}{n} + \frac{\beta \eta }{\alpha^2} j , \label{eq:geneOhm}
\end{eqnarray}
where we highlight the momentum conservation eq. (\ref{eq:momentum}), the current equation (\ref{eq:current}) and generalized Ohm's law (\ref{eq:geneOhm}). Note that the $T \, \nabla \mbox{ln} \, n $ term at the right-hand side of (\ref{eq:geneOhm}) has no contribution since the curl operator in the Faraday equation cancels it. However, this cancellation is no more true in the general gas dynamics case because $\frac{\nabla p_e}{n}$ may not be a gradient in general, where $p_e$ is the electron pressure. 

In all what follows, we assume $\alpha^2 \eta = 1$ in order to keep the Lorentz force term in (\ref{eq:momentum}) of order $1$ in all the various scalings below. Then, eqs (\ref{eq:momentum}),  (\ref{eq:current}) and  (\ref{eq:geneOhm}) are written :
\begin{eqnarray}
& & \hspace{-1cm}  \partial_t (n  u )  +    \nabla ( n   u  \otimes u  ) + \nabla (2 n T)  =  j \times B ,  \label{eq:momentum2} \\
& & \hspace{-1cm} \frac{1}{\alpha^2} j =  n (u - u_e), \label{eq:current2} \\
& & \hspace{-1cm} E + u \times B =  \frac{1}{\alpha^2} \left[ - T  \nabla ( \mbox{ln} \, n ) +  \frac{j \times B}{n} \right] + \frac{\beta }{\alpha^4} j  , \label{eq:geneOhm2}
\end{eqnarray}
the other equations being unchanged. There are only two dimensionless parameters left: $\frac{1}{\alpha^2}$ and $\frac{\beta }{\alpha^4}$ and they only appear in (\ref{eq:current2}) and in (\ref{eq:geneOhm2}). So, the various types of MHD model correspond to the various choices of scalings for these two parameters. In particular, we have 
\begin{enumerate}
\item If both $\frac{1}{\alpha^2} \to 0$  and  $\frac{\beta }{\alpha^4} \to 0 $, then the generalized Ohm's law reduces to the standard ideal Ohm's law while the electron and ion velocities become identical:
$$ E + u \times B =  0 , \quad u_e = u .  $$
This yields ideal MHD. 
\item If $\frac{1}{\alpha^2} \to 0$ but $\frac{\beta }{\alpha^4} \to 1$, then the resistive term in the generalized Ohm's law is kept but the electron and ion velocities are still identical: 
$$ E + u \times B = j , \quad u_e = u .  $$
This gives rise to resistive MHD. 
\item If $\frac{1}{\alpha^2} \to 1$ but $\frac{\beta }{\alpha^4} \to 0$, then, the ion and electron velocities differ and additionally, the generalized Ohm's law has the form: 
$$ E + u \times B =   -  T \, \nabla ( \mbox{ln} \,  n )  +  \frac{j \times B}{n} . $$
As already mentioned, the first term at the right-hand side has no contribution. The second one is the Hall term. This gives rise to the  Hall MHD. 
\item Finally, if both $\frac{1}{\alpha^2} \to 1$, $\frac{\beta }{\alpha^4} \to 1$, then, the ion and electron velocities differ and both the Hall and resistive terms appear. 
$$ E + u \times B =   - T \,   \nabla ( \mbox{ln} n )  +  \frac{j \times B}{n} + j . $$
\end{enumerate}

Our study takes place in the context of the last regime, where both the resistive and Hall terms are equally important. Additionaly, we assume incompressible viscous fluid motion. In this case, the Hall MHD system can be written according to (\ref{eq:dtu})-(\ref{eq:divB}). 

We note that it is easy to extend this system to the viscous isentropic resistive compressible Hall MHD as follows (assuming all the physical constants equal to $1$ except the viscosity here denoted by $\nu$): 
\begin{eqnarray}
& & \hspace{-1cm} \partial_t n +   \nabla \cdot (n u ) = 0 ,  \label{eq:CHMHD_mass}\\
& & \hspace{-1cm}  \partial_t (n  u )  +    \nabla ( n   u  \otimes u  ) + \nabla p(n)  =  (\nabla \times B) \times B + \nu \nabla \cdot (\nabla u + (\nabla u)^T),   \label{eq:CHMHD_momentum} \\
& & \hspace{-1cm} \partial_t B +  \nabla \times \left( 
B \times u + \frac{(\nabla \times B)  \times B}{n} \right) = - \nabla \times (\nabla \times B) ,    \label{eq:CHMHD_faraday}
\end{eqnarray}

For this system, we have the following magneto-helicity conservation relation: 
\begin{eqnarray}
& & \hspace{-1cm} 
\frac{d}{dt} \int_{{\mathbb R}^3} B \cdot A \, dx + 2 \int_{{\mathbb R}^3} B \cdot (\nabla \times B) \, dx = 0,     \label{eq:CHMHD_helicity}
\end{eqnarray}
where $A$ such that $B = \nabla \times A$ is any vector potential of $B$.  
To prove this relation, we note, using Green's formula, that 
\begin{eqnarray*}
& & \hspace{-1cm} 
\frac{d}{dt} \int_{{\mathbb R}^3} B \cdot A \, dx = 2 \int_{{\mathbb R}^3} A \cdot B_t \, dx .     \end{eqnarray*}
Then, taking the scalar product of (\ref{eq:CHMHD_faraday}) with $2A$ easily gives the result. 

\begin{remark}
The viscosity term at the right-hand side of (\ref{eq:CHMHD_momentum}) involves the rate of strain tensor $\sigma(u) = \nabla u + (\nabla u)^T$. In usual gas dynamics, the viscosity term involves the traceless rate of strain tensor $\sigma_0(u) = \nabla u + (\nabla u)^T - (2/3) (\nabla \cdot u) \mbox{Id}$. However, a careful computation of the viscosity in the case of isothermal gas dynamics shows that, in this case, the right tensor is the full rate-of-strain tensor $\sigma(u)$ and not its trace-free counterpart $\sigma_0(u)$. 
\end{remark}

%%%%%%%%%%%%%%%%%%%%%%%%%%%%%%%%%%%%%%%%%%%%%%%%%%%%%%%%%%%%%%%%%%%%%%%%%%%%%%%%%%%%%%%%%%%%%%%%
%%%%%%%%%%%%%%%%%%%%%%%%%%%%%%%%%%%%%%%%%%%%%%%%%%%%%%%%%%%%%%%%%%%%%%%%%%%%%%%%%%%%%%%%%%%%%%%%
%%%%%%%%%%%%%%%%%%%%%%%%%%%%%%%%%%%%%%%%%%%%%%%%%%%%%%%%%%%%%%%%%%%%%%%%%%%%%%%%%%%%%%%%%%%%%%%%
%%%%%%%%%%%%%%%%%%%%%%%%%%%%%%%%%%%%%%%%%%%%%%%%%%%%%%%%%%%%%%%%%%%%%%%%%%%%%%%%%%%%%%%%%%%%%%%%
\setcounter{equation}{0}
\section{Derivation of the Hall-MHD equations from kinetic equations}
\label{sec:derivation_kinetic}

In this section, we provide a kinetic formulation of the Hall MHD problem. We start from a kinetic equation for the ion distribution function $f(x,v,t)$ of the plasma, where $x$ is the position, $v$ the velocity and $t$ the time. This distribution function is a solution of the following kinetic equation
\begin{eqnarray}
& & \hspace{-1cm} \partial_t f + v \cdot \nabla_x f + \frac{e}{m} (E + v \times B) \cdot \nabla_v f = Q(f) \, , \label{Vlasov} 
\end{eqnarray}
where  $e$ is the positive ion charge, supposed equal to the absolute value of the elementary charge, $m$ is their mass, $E(x,t)$ and $B(x,t)$ are the electric and magnetic fields, and $Q(f)$ is the collision operator for electron-ion collisions. We respectively introduce the ion density $n$, mean velocity $u$ and energy $W$ by 
\begin{eqnarray*}
& & \hspace{-1cm} n = \int f \, dv \, , \quad n u =  \int f \, v \, dv \, , \quad W =  \int f \, m \frac{|v|^2}{2} \, dv, 
\end{eqnarray*}
and the temperature by 
\begin{eqnarray*}
& & \hspace{-1cm} \frac{3}{2} n T = W - \frac{1}{2} n |u|^2 = \int f \, m \frac{|v-u|^2}{2} \, dv. 
\end{eqnarray*}

We assume that the electrons are described by their fluid quantities, namely their density $n_e(x,t)$, their fluid velocity $u_e(x,t)$ and their temperature $T_e(x,t)$. The use of a fluid model for the electrons while the ions are treated kinetically can be justified by the small electron to ion mass ratio. A formally rigorous justification of this can be found e.g. in \cite{Deg_Luc}. The electron density is supposed equal to the ion density by quasineutrality:
\begin{eqnarray}
& & \hspace{-1cm} n_e = n \, .\label{quasineutralite} 
\end{eqnarray}
The electron momentum conservation equation, when the transport term is neglected due to their small mass, gives rise to the generalized Ohm's law (see (\ref{eq:electron_momentum}) where all terms in factor of $m_e$ are set to zero):
\begin{eqnarray}
& & \hspace{-1cm} \nabla_x (n T_e) + e n (E + u_e \times B) = e \eta n j 
\, , \label{Ohm} 
\end{eqnarray}
where $m_e$ is the electron mass, $j$ is the current density and $\eta$ is the resistivity. The current density is given by:
\begin{eqnarray}
& & \hspace{-1cm} j = e n (u - u_e) 
\, .
\label{courant} 
\end{eqnarray}
Since the electron mass is neglected and assuming monoatomic gas equation of state, the electron energy $W_e$ can be expressed in terms of the electron temperature $T_e$ by
$$ W_e = \frac{3}{2} n T_e.  $$
For the same reason, the mass is neglected in the electron energy flux, which reads $ \frac{5}{2} n T_e u_e$. 

For simplicity, we consider a model electron-ion collision operator as follows: 
\begin{eqnarray}
& & \hspace{-1cm} Q(f) = \frac{e^2 \eta n}{m} \nabla_v \cdot ((v-u_e) f + \frac{T_e}{m} \nabla_v f) \, . 
\label{coll} 
\end{eqnarray}
The first term expresses the relaxation of the ion velocity to the electron one, while the second one expresses the relaxation of the ion temperature to the electron one. More realistic expressions of the electron-ion collision operator can be found in the literature (see e.g. \cite{Braginskii}), but this model is chosen for the sake of simplicity of exposition. The rate of change of the ion momentum is given by: 
$$ \int Q(f) \, m v \, dv = - e \eta n j , $$
and is the opposite of the right-hand side of (\ref{Ohm}), which is consistent with the total momentum conservation of the electron-ion collisions. The rate of change of the ion energy is given by 
$$ \int Q(f) \, \frac{m |v|^2}{2} \, dv = 
- e \eta n j \cdot u + 3 \frac{e^2 \eta n^2}{m} (T_e - T).$$
By total energy conservation in electron-ion collisions, the rate of change of the electron total energy is the opposite.  Then, the electron energy conservation equation reads: 
\begin{eqnarray}
& & \hspace{-1cm}  \partial_t \left( \frac{3}{2} n T_e \right) + \nabla_x \cdot  \left( \frac{5}{2} n T_e u_e\right) = - e n u_e \cdot E + e \eta n j \cdot u + 3 \frac{e^2 \eta n^2}{m} (T - T_e) . \label{eq:electron_energy_bal} 
\end{eqnarray}
In this equation, the first term of the right-hand side is the work done by the electrons in the Lorentz force, while the last two terms are due to the electron-ion collisions. Taking the scalar product of (\ref{Ohm}) by $u_e$ and subtracting it to (\ref{eq:electron_energy_bal}) leads to 
\begin{eqnarray}
& & \hspace{-1cm}  (\partial_t + u_e \cdot \nabla_x ) \left( \frac{3}{2} n T_e \right) +  \frac{5}{2} n T_e \nabla_x \cdot u_e = \eta |j|^2 + 3 \frac{e^2 \eta n^2}{m} (T - T_e) . \label{eq:electron_temp_bal} 
\end{eqnarray}
The first term of the right-hand side is Joule heating of the electrons, while the second term is the electron temperature relaxation to the ion temperature. 

The magnetic field evolves according to the Faraday equation
\begin{eqnarray}
& & \hspace{-1cm} \partial_t B + \nabla_x \times E = 0  
\, , 
\label{faraday} 
\end{eqnarray}
and the current is linked to the magnetic field by Ampere's law: 
\begin{eqnarray}
& & \hspace{-1cm} \nabla_x \times B = \mu_0 j  
\, ,
\label{ampere} 
\end{eqnarray}
As in the previous section, the displacement current has been neglected. 

As a summary, the considered kinetic model is as follows: 
\begin{eqnarray}
& & \hspace{-1cm} \partial_t f + v \cdot \nabla_x f + \frac{e}{m} (E + v \times B) \cdot \nabla_v f  = \frac{e^2 \eta n}{m} \nabla_v \cdot ((v-u_e) f + \frac{T_e}{m} \nabla_v f) \, , \label{Vlasov_0} \\
& & \hspace{-1cm} \nabla_x (n T_e) + e n (E + u_e \times B) = e \eta n j 
\, , \label{Ohm_0} \\ 
& & \hspace{-1cm} (\partial_t + u_e \cdot \nabla_x ) \left( \frac{3}{2} n T_e \right) +  \frac{5}{2} n T_e \nabla_x \cdot u_e = \eta |j|^2 + 3 \frac{e^2 \eta n^2}{m} (T - T_e) \, , \label{energie_0} \\ 
& & \hspace{-1cm} \partial_t B + \nabla_x \times E = 0  
\, . 
\label{faraday_0} \\ 
& & \hspace{-1cm} \nabla_x \times B = \mu_0 j  
\, ,
\label{ampere_0} \\ 
& & \hspace{-1cm}  j = e n (u - u_e) 
\, , 
\label{courant_0} \\
& & \hspace{-1cm} n = \int f \, dv \, , \quad n u =  \int f \, v \, dv \, , \quad \frac{3}{2} n T = \int f \, m \frac{|v-u|^2}{2} \, dv. \label{moment_0}
\end{eqnarray}

Now, we link this system to Hall-MHD by taking the moments of the ion kinetic equation. Integrating (\ref{Vlasov_0}) with respect to $v$ after premultiplying it successively by $1$, $m v$ or $m \frac{|v|^2}{2}$, we get the following ion mass, momentum and energy balance equations: 
\begin{eqnarray}
& & \hspace{-1cm} \partial_t n + \nabla_x \cdot (nu)  = 0 \, , \label{cons_dens} \\
& & \hspace{-1cm} m \left( \partial_t (nu) + \nabla_x \cdot (nu \otimes u) \right) + \nabla_x \cdot {\mathbb P} = e n (E + u \times B) - e \eta n j  \, , \label{cons_moment} \\
& & \hspace{-1cm} \partial_t W + \nabla_x \cdot (W u + {\mathbb P} u + {\mathbf q})   = e n E \cdot u - e \eta n j \cdot u + 3 \frac{e^2 \eta n^2}{m} (T_e - T)
\, , \label{cons_energie} 
\end{eqnarray}
where ${\mathbb P}$ and ${\mathbf q}$ are the stress tensor and heat flux vector, given by:  
\begin{eqnarray}
& & \hspace{-1cm} {\mathbb P} = m \int f \, (v-u) \otimes (v-u) \, dv, \quad {\mathbf q} = \frac{m}{2} \int f \, (v-u) |v-u|^2 \, dv\, . \label{mom_ord_sup} 
\end{eqnarray}

By combining (\ref{cons_moment}) with the generalized Ohm law (\ref{Ohm_0}), we obtain the total fluid momentum balance
\begin{eqnarray}
& & \hspace{-1cm} m \left( \frac{\partial}{\partial t} (nu) + \nabla_x \cdot (nu \otimes u) \right) + \nabla_x \cdot ({\mathbb P} + n T_e \mbox{Id})  = j \times B  \, . \label{cons_moment_fluide} 
\end{eqnarray}
Using Ampere's equation (\ref{ampere_0}), the fact that $\nabla_x \cdot B = 0$ and the vector identity $(\nabla_x \times B) \times B = \nabla_x \cdot (B \otimes B) - \nabla_x (|B|^2/2)$,  the total fluid momentum balance can be written in conservative form:  
\begin{eqnarray}
& & \hspace{-1cm} \frac{\partial}{\partial t} (m n u) + \nabla_x \cdot (m nu \otimes u  - \frac{1}{\mu_0} B \otimes B +  {\mathbb P}_{\mbox{\scriptsize tot}} ) = 0  \, , \label{cons_moment_tot} 
\end{eqnarray}
where the total pressure tensor ${\mathbb P}_{\mbox{\scriptsize tot}}$ is written
\begin{eqnarray}
& & \hspace{-1cm} {\mathbb P}_{\mbox{\scriptsize tot}}  = {\mathbb P} + (n T_e  +  \frac{|B|^2 }{2 \mu_0} ) \mbox{Id}   \, . \label{pression_tot} 
\end{eqnarray}
Similarly, by adding the energy conservation equations (\ref{eq:electron_energy_bal} ) and (\ref{cons_energie}), we get, for the total fluid energy $W_f = W + W_e$:
\begin{eqnarray}
& & \hspace{-1cm} \frac{\partial}{\partial t} W_f + \nabla_x \cdot ( (W_e + p_e) u_e + 
W u + {\mathbb P} u + {\mathbf q})   = E \cdot j \, . \label{energie_fluide} 
\end{eqnarray}
The Faraday equation (\ref{faraday_0}) implies that 
\begin{eqnarray}
& & \hspace{-1cm} \frac{\partial }{\partial t} \left( \frac{|B|^2}{2 \mu_0} \right) + \frac{1}{\mu_0} B \cdot (\nabla_x \times E) = 0  
\, . 
\label{energie_B} 
\end{eqnarray}
By adding (\ref{energie_fluide}) and (\ref{energie_B}) and using Ampere's law (\ref{ampere_0}), the total energy (which is the sum of the total fluid energy and the magnetic energy) $W_{\mbox{\scriptsize tot}}= W_f +\frac{|B|^2}{2 \mu_0}$ satisfies the following conservation law: 
\begin{eqnarray}
& & \hspace{-1cm} \frac{\partial}{\partial t} W_{\mbox{\scriptsize tot}} + \nabla_x \cdot ( (W_e + p_e) u_e + W u + {\mathbb P} u + {\mathbf q} + \frac{1}{\mu_0} E \times B)   = 0 \, , \label{energie_totale} 
\end{eqnarray}

The MHD equations are obtained through the closure assumptions that ${\mathbb P} = n T \mbox{Id}$, ${\mathbf q}=0$, which can be justified e.g. by a Maxwellian closure, i.e. assuming that $f = M_{n,u, T}$ with 
$$ M_{n,u, T} = \frac{n}{\left( \frac{2 \pi T}{m} \right)^{3/2}} \exp \left( - \frac{m |v-u|^2}{2 T} \right). $$
The Maxwellian closure itself can be justified if ion-ion collisions are strong enough to relax the distribution $f$ quickly to $M_{n,u, T}$. However, in many instances, the Maxwellian closure is used in spite of not being fully justified. In this case, we obtain the 2-temperature, Hall, resistive compressible MHD equations which are as follows: 
\begin{eqnarray}
& & \hspace{-1cm} \partial_t n + \nabla_x \cdot (nu)  = 0 \, , \label{cons_dens_2} \\
& & \hspace{-1cm} \partial_t (m n u) + \nabla_x \cdot (m nu \otimes u  - \frac{1}{\mu_0} B \otimes B +  ( n(T_e + T)  +  \frac{|B|^2 }{2 \mu_0} ) \mbox{Id} ) = 0  \, , \label{cons_moment_tot_2}  \\
& & \hspace{-1cm} \partial_t \left( W + \frac{3}{2} n T_e + \frac{|B|^2}{2 \mu_0} \right)  + \nabla_x \cdot (  W u + n T u + \frac{5}{2} n T_e u_e + \frac{1}{\mu_0} E \times B)   = 0 \, , \label{energie_totale_2} \\
& & \hspace{-1cm} \partial_t B + \nabla_x \times E = 0  
\, , 
\label{faraday_2} \\
& & \hspace{-1cm} \nabla_x (n T_e) + e n (E + u_e \times B) = e \eta n j 
\, , \label{Ohm_2} \\ 
& & \hspace{-1cm} (\partial_t + u_e \cdot \nabla_x ) \left( \frac{3}{2} n T_e \right) +  \frac{5}{2} n T_e \nabla_x \cdot u_e = \eta |j|^2 + 3 \frac{e^2 \eta n^2}{m} (T - T_e) \, , \label{energie_2} \\  
& & \hspace{-1cm} \nabla_x \times B = \mu_0 j  
\, ,
\label{ampere_2} \\ 
& & \hspace{-1cm}  j = e n (u - u_e) .
\label{current_2}
\end{eqnarray}
The first four equations are the basic conservation laws of mass, total momentum, total energy and magnetic field. Eq. (\ref{Ohm_2}) is the generalized Ohm law and provides the expression for $E$. Eq. (\ref{energie_2}) provides the evolution equation for $T_e$. Finally, Ampere's eq. (\ref{ampere_2}) defines $j$ and (\ref{current_2}) defines $u_e$. The fact that $u_e \not = u$ gives rise to the Hall term. Now, if $T = T_e$, then, eq. (\ref{energie_2}) can be removed and one gets the single temperature Hall resistive compressible MHD equations. Another simplification is to suppose that the common ion and electron temperatures are constant (isothermal assumption). In this case, the total energy equation (\ref{energie_totale_2}) is a consequence of the Faraday and momentum eqs. (\ref{faraday_2}), (\ref{cons_moment_tot_2}) and can be removed. Then, we find the model of section \ref{sec:derivation}. 

A kinetic formulation of the model considered in section \ref{sec:derivation} is also obtained from the kinetic model (\ref{Vlasov_0})-(\ref{moment_0}) by supposing that the electron and ion temperatures are the same constant $T$. This kinetic model is written below: 
\begin{eqnarray}
& & \hspace{-1cm} \partial_t f + v \cdot \nabla_x f + \frac{e}{m} (E + v \times B) \cdot \nabla_v f  = \frac{e^2 \eta n}{m} \nabla_v \cdot ((v-u_e) f + \frac{T}{m} \nabla_v f) \, , \label{eq:kin_isoth_fi} \\
& & \hspace{-1cm} T \nabla_x n + e n (E + u_e \times B) = e \eta n j 
\, , \label{eq:kin_isoth_Ohm} \\ 
& & \hspace{-1cm} \partial_t B + \nabla_x \times E = 0  
\, ,  \label{eq:kin_isoth_Faraday}\\ 
& & \hspace{-1cm} \nabla_x \times B = \mu_0 j  
\, , \label{eq:kin_isoth_Ampere}\\ 
& & \hspace{-1cm}  j = e n (u - u_e) 
\, . \label{eq:kin_isoth_j}
\end{eqnarray}
By eliminating $E$, $j$ and $u_e$ respectively using (\ref{eq:kin_isoth_Ohm}), (\ref{eq:kin_isoth_Ampere}), (\ref{eq:kin_isoth_j}), and after some easy algebra, we find the following coupled Fokker-Planck Faraday system for $f$ and $B$: 
\begin{eqnarray}
& & \hspace{-1cm} \partial_t f + v \cdot \nabla_x f + \frac{e}{m} \left[ (v-u) \times B + \frac{1}{\mu_0 e n} (\nabla_x \times B) \times B - \frac{T}{e} \nabla_x \mbox{ln} \, n \right] \cdot \nabla_v f  = \nonumber \\
& & \hspace{6cm}
= \frac{e^2 \eta n}{m} \nabla_v \cdot ((v-u) f + \frac{T}{m} \nabla_v f) \, , \label{eq:kin_isoth_fi2} \\
& & \hspace{-1cm} \partial_t B + \nabla_x \times \left( B \times u +  \frac{1}{\mu_0 e n} (\nabla_x \times B) \times B \right) = - \frac{\eta}{\mu_0} \nabla_x \times (\nabla_x \times B)
\, .  \label{eq:kin_isoth_Faraday2}
\end{eqnarray}
For the kinetic eq. (\ref{eq:kin_isoth_fi2}), we have the following entropy dissipation identity associated to the entropy $H(f)$: 
\begin{eqnarray*}
& & \hspace{-1cm} \frac{dH(f)}{dt} + \frac{e^2 \eta}{T} \int_{(x,v) \in {\mathbb R}^3 \times {\mathbb R}^3} n \, \frac{ \left| (v-u) f + \frac{T}{m} \nabla_v f \right|^2}{f} \, dx \, dv = 0,  \\
& & \hspace{-1cm}
H(f)= \int_{(x,v) \in {\mathbb R}^3 \times {\mathbb R}^3} f \, \mbox{ln} \, f \, dx \, dv. 
\end{eqnarray*}
Particle-in-Cell simulations of this model can be found in \cite{D3GST}. Now, taking the first two moments of (\ref{eq:kin_isoth_fi2}), we get the isothermal resistive Hall MHD model which was written in dimensionless form at (\ref{eq:CHMHD_mass})-(\ref{eq:CHMHD_faraday}) (taking $\nu = 0$ and $p(n) = 2 T n$). It is interesting to note that in order to get a kinetic model for the standard resistive isothermal MHD equations, we need to neglect the Hall term $\frac{1}{\mu_0 e n} (\nabla_x \times B) \times B$ in the Faraday eq. (\ref{eq:kin_isoth_Faraday2}), but we must retain the corresponding term in the kinetic equation (\ref{eq:kin_isoth_fi2}). Therefore, a kinetic formulation for standard resistive isothermal MHD is as follows: 
\begin{eqnarray}
& & \hspace{-1cm} \partial_t f + v \cdot \nabla_x f + \frac{e}{m} \left[ (v-u) \times B + \frac{1}{\mu_0 e n} (\nabla_x \times B) \times B - \frac{T}{e} \nabla_x \mbox{ln} \, n \right] \cdot \nabla_v f  = \nonumber \\
& & \hspace{6cm}
= \frac{e^2 \eta n}{m} \nabla_v \cdot ((v-u) f + \frac{T}{m} \nabla_v f) \, , \label{eq:kin_isoth_fi3}\\
& & \hspace{-1cm} \partial_t B + \nabla_x \times \left( B \times u \right) = - \frac{\eta}{\mu_0} \nabla_x \times (\nabla_x \times B) 
\, . \label{eq:kin_isoth_Faraday3}
\end{eqnarray}
We can also get a kinetic formulation for the ideal isothermal MHD equations by neglecting the resistive term 
$\frac{\eta}{\mu_0} \nabla_x \times (\nabla_x \times B)$ at the right-hand side of the Faraday eq. (\ref{eq:kin_isoth_Faraday3}). The collision term at the right-hand side of (\ref{eq:kin_isoth_fi3}) has no contribution in the mass and momentum conservation equations and can be either kept or removed without modifying the corresponding balance equations.

%%%%%%%%%%%%%%%%%%%%%%%%%%%%%%%%%%%%%%%%%%%%%%%%%%%%%%%%%%%%%%%%%%%%%%%%%%%%%%%%%%%%%%%%%%%%%%%%
%%%%%%%%%%%%%%%%%%%%%%%%%%%%%%%%%%%%%%%%%%%%%%%%%%%%%%%%%%%%%%%%%%%%%%%%%%%%%%%%%%%%%%%%%%%%%%%%
%%%%%%%%%%%%%%%%%%%%%%%%%%%%%%%%%%%%%%%%%%%%%%%%%%%%%%%%%%%%%%%%%%%%%%%%%%%%%%%%%%%%%%%%%%%%%%%%
%%%%%%%%%%%%%%%%%%%%%%%%%%%%%%%%%%%%%%%%%%%%%%%%%%%%%%%%%%%%%%%%%%%%%%%%%%%%%%%%%%%%%%%%%%%%%%%%
\setcounter{equation}{0}
\section{Existence result for the incompressible, viscous, resistive Hall-MHD}
\label{sec:existence}

In the paper, we use the function spaces 
\begin{eqnarray*} 
& & \hspace{-1cm}  H_{per}(\Omega )=\lbrace B\: \in \: (H^1(\Omega ))^3 \, |  \, \nabla \cdot B=0 \text{ on } \Omega \text{  with periodicity conditions} \rbrace . 
\end{eqnarray*}
We use $\langle A,B\rangle =\int_\Omega A\cdot B \, dx$, for any pair $A,B \in(L^2(\Omega))^3$. 
We remark that
\begin{equation}
\left\|\nabla B \right\|_{(L^2(\Omega ))^9} =\left\|\nabla \times B \right\|_{(L^2(\Omega ))^3} . 
\label{defnorme}
\end{equation}
To prove theorem \ref{th:global}, we just focus on the Hall problem itself. The extension to the full Hall-MHD problem is explained at the end of the section. So, our goal is now to prove the following existence theorem for the Hall problem. We introduce the following weak formulations of the Hall MHD problem: 

\medskip
\noindent
Find $B\in L^\infty (0,T, L^2(\Omega))$ $ \cap $ $ L^2((0,T),$ $H_{per}(\Omega))$ such that for any $A\in H_{per}(\Omega)$:
\begin{equation}
\left\langle A,\partial_t B \right\rangle+\left\langle \nabla \times A,(\nabla \times B)\times B \right\rangle+\left\langle \nabla \times A,\nabla \times B \right\rangle =0.
\label{fvperiodique}
\end{equation}

\begin{theorem}
Assume that $B_0 \in (L^2(\Omega ))^3$ with $\nabla \cdot B_0=0$. Then, there exists a global weak solution $B$ for the Hall problem (\ref{eq:dtB2}), (\ref{eq:divB2}).  Moreover, we have $B \in L^{\infty }((0,T),L^2(\Omega))$ $ \cap $ $ L^2((0,T),$ $H_{per}(\Omega))$ and $\partial_t B \in L^{\frac{4}{3}}((0,T),H^{-2}(\Omega))$.
\label{thm}
\end{theorem}

The proof is based on the construction of an approximate solution by Galerkin's method. Uniform a priori bounds on these approximate solutions will allow us to pass to the limit thanks to standard compactness arguments. 

We define the Fourier basis $(\phi_k)_{k \in {\mathbb Z}^3}$, with $\phi_k = e^{2 \pi i k \cdot x}$. We denote by $|k|= |k_1|+|k_2|+|k_3|$.  The Fourier basis provides a complete ortho-normal basis of $L^2(\Omega)$. We denote by $P^N$ the projection onto the Fourier basis with index $|i|\leq N$.
We define an approximate solution  $B^N:(0,T)\mapsto H_{per} (\Omega )$ of problem (\ref{eq:dtu}), (\ref{eq:dtB}) of the form:
\begin{equation}
B^N(t)=\sum_{i \in {\mathbb Z}^3, \, |i|\leq N} B^N_i(t) \varphi_i 
\label{decompo}
\end{equation}
with the divergence free constraint $i \cdot B^N_i(t)= 0$, satisfying
\begin{eqnarray}
& & \hspace{-1cm} \left\langle A^N ,\partial_t B^N(t) \right\rangle
+\left\langle \nabla \times A^N, (\nabla \times B^N(t))\times B^N(t) \right\rangle
+\left\langle \nabla \times A^N, \nabla \times B^N(t) \right\rangle = 0, \label{projection}
\end{eqnarray}
for all $A^N$ of the form 
\begin{equation*}
A^N=\sum_{i=1}^N A^N_i \varphi_i , \quad i \cdot A^N_i= 0,
\end{equation*}
and with initial condition $B^N(0) = P^N B_0$. 

\begin{lemma}\label{borne}
There exists a global in time solution $B^N(t)$ of (\ref{projection}) which is uniformly (independently of $N$) bounded in $L^{\infty }((0,T),L^2(\Omega ))$~$ \cap $~$ L^2((0,T),H^1(\Omega ))$ such that $\partial_t B^N$ is uniformly bounded in $L^{\frac{4}{3}}((0,T),H^{-2}(\Omega))$.
\end{lemma}

{\bf Proof.} Throughout of the proof, $C$ denotes a generic constant. We take $A=B^N$ in (\ref{projection}) and get:
\begin{equation}
\left\langle B^N,\partial_t B^N \right\rangle+\Vert \nabla \times B^N \Vert^2_{(L^2(\Omega ))^3}=0 .
\label{30}
\end{equation}
Thus, we have:
\begin{equation}
 \Vert B^N(t) \Vert ^2_{L^2(\Omega )}+2\int_0^t  \Vert \nabla B^N(s) \Vert ^2_{(L^2(\Omega ))^9} \, ds =\Vert B^N_0\Vert ^2_{L^2(\Omega )} .
 \label{eq:BNsquare}
\end{equation}
This shows the uniform bound of $B^N$ in $L^{\infty }((0,T),L^2(\Omega ))$~$ \cap $~$ L^2((0,T),H^1(\Omega ))$.

To show the time regularity, we take $A \in H^2(\Omega)^3$, such that $\nabla \cdot A = 0$ and take $A^N = P^N A$. 
\begin{equation}
\left\langle A,\partial_t B^N \right\rangle =
\left\langle A^N,\partial_t B^N \right\rangle =
-\left\langle \nabla \times A^N,(\nabla \times B^N)\times B^N \right\rangle
-\left\langle \nabla \times A^N,\nabla \times B^N\right\rangle. \label{eq:AtBN}
\end{equation}
We recall that 
\begin{eqnarray}
& & \hspace{-1cm} (\nabla \times B)\times B = \nabla \cdot(B \otimes B)-\nabla \left( \tfrac{\left| B \right|^2}{2}\right). 
\label{eg1} 
\end{eqnarray}
Introducing (\ref{eg1}) into (\ref{eq:AtBN}) and noting that, by Green's formula, a gradient and a curl are orthogonal in $L^2$, we find 
\begin{eqnarray}
& & \hspace{-1cm} \left\langle A,\partial_t B^N \right\rangle
=
\left\langle \nabla \times A^N,\nabla \cdot (B^N \otimes B^N)\right\rangle 
+  \left\langle \nabla \times A^N,\nabla \times B^N\right\rangle. \label{eq:AtBN_2}
\end{eqnarray}

To estimate the first term at the right-hand side of (\ref{eq:AtBN_2}), we use Green's formula and get: 
$$ \left\langle \nabla \times A^N,\nabla \cdot (B^N \otimes B^N)\right\rangle  = -  \left\langle B^N,(B^N\cdot \nabla ) \nabla \times A^N \right\rangle. $$
Using H\"older's inequality, and remarking that $\Vert A^N \Vert_{H^2(\Omega )} \leq \Vert A \Vert_{H^2(\Omega )}$, we get 
\begin{eqnarray}
& & \hspace{-1cm} 
\vert \left\langle B^N,(B^N\cdot \nabla ) \nabla \times A^N\right\rangle \vert 
\leq C \Vert B^N \Vert _{L^4(\Omega )}^2 \Vert A^N \Vert _{H^2(\Omega )} \leq C \Vert B^N \Vert _{L^4(\Omega )}^2 \Vert A \Vert _{H^2(\Omega )} . 
\label{part1}
\end{eqnarray} 
By Gagliardo-Nirenberg's inequality \cite{Evans},
\begin{equation*}
\Vert B^N \Vert _{L^4(\Omega )}^2\leq C \Vert \nabla B^N \Vert _{L^2(\Omega )}^\frac{3}{2} \, (\Vert B^N \Vert _{L^2(\Omega )}^\frac{1}{2} +1) ,
\end{equation*} 
and using the uniform $L^2$ bound, we can write:
\begin{equation*}
\vert \left\langle B^N,(B^N\cdot \nabla ) \nabla \times A^N \right\rangle \vert \leq C \Vert \nabla B^N \Vert _{L^2(\Omega )}^\frac{3}{2} \Vert A\Vert _{H^2(\Omega )} .
\end{equation*} 
The second term is simply estimated by using Cauchy-Schwartz inequality:
\begin{equation*} 
\vert \left\langle \nabla \times A^N,\nabla \times B^N\right\rangle \vert \leq C \, \Vert \nabla B^N \Vert _{L^2(\Omega )} \, \Vert A^N\Vert _{H^1(\Omega )} \leq C \, \Vert \nabla B^N \Vert _{L^2(\Omega )} \, \Vert A \Vert _{H^1(\Omega )}
\end{equation*} 
Collecting these estimates, we  obtain:
\begin{eqnarray*} 
\vert \left\langle A,\partial_t B^N \right\rangle \vert 
& \leq & C\left[ \Vert \nabla B^N \Vert _{L^2(\Omega )}^\frac{3}{2} \, \Vert A\Vert _{H^2(\Omega )} 
+ \Vert \nabla B^N \Vert _{L^2(\Omega )} \, \Vert A\Vert _{H^1(\Omega )}\right] \\
&\leq & C \left[ \Vert \nabla B^N \Vert _{L^2(\Omega )}^\frac{3}{2} + 1 \right] \Vert A\Vert _{H^2(\Omega )} .
\end{eqnarray*} 
Therefore,  
\begin{equation*} 
\Vert \partial_t B^N\Vert _{H^{-2}(\Omega )}\leq   C \left[ \Vert \nabla B^N \Vert _{L^2(\Omega )}^\frac{3}{2} + 1 \right] . 
\end{equation*} 
Thus:
\begin{equation*} 
\Vert \partial_t B^N\Vert _{H^{-2}(\Omega )}^\frac{4}{3}\leq  C \left[ \Vert \nabla B^N \Vert _{L^2(\Omega )}^2 + 1 \right] . 
\end{equation*} 
Thanks to (\ref{eq:BNsquare}), the right-hand side is integrable on $(0,T)$ and we get
\begin{equation}
\int_0^{T} \Vert \partial_t B^N\Vert _{H^{-2}(\Omega )}^\frac{4}{3}dt \leq C ,
\label{ineg*}
\end{equation}
which ends the proof of Lemma \ref{borne}. \endproof

\noindent
Thanks to this Lemma, we can proceed to the 

\medskip
\noindent
{\bf Proof of Theorem \ref{thm}.} According to Lemma \ref{borne}, the sequence $( B^N )_{N \in \mathbb{N}}$ is uniformly bounded in $L^2((0,T),H^1(\Omega ))$ with $(\partial_tB^N)_{N \in \mathbb{N}}$ uniformly bounded in $L^{\frac{4}{3}}((0,T),H^{-2}(\Omega))$. Consequently, by virtue of Lions-Aubin Lemma \cite{J.L.Lions}, $(B^N )_{N \in \mathbb{N}}$ is compact in $L^2((0,T),L^2(\Omega ))$. Therefore, there exists a subsequence $( B^{N_k} )_{k \in \mathbb{N}}$ and a function $B$ in $L^2((0,T),H^1(\Omega ))$ with $\partial_tB$ in $L^{\frac{4}{3}}((0,T),H^{-2}(\Omega ))$, such that:
\begin{equation}
\left\{ 
\begin{aligned}
&B^{N_k}\rightharpoonup B &\quad &\text{weak star in } L^{\infty }((0,T),L^2(\Omega ))\\
&B^{N_k}\rightharpoonup B &\quad &\text{weakly in } L^2((0,T),H^1(\Omega ))\\
&B^{N_k}\rightarrow B &\quad &\text{strongly in }L^2((0,T),L^2(\Omega ))\\
&\partial_tB^{N_k}\rightharpoonup \partial_tB &\quad &\text{weakly in } L^{\frac{4}{3}}((0,T),H^{-2}(\Omega))\\
\end{aligned}
\right.
\label{cvfaible}
\end{equation}

We take $A \in H^3(\Omega)^3$, such that $\nabla \cdot A = 0$, take $A^{N_k} = P^{N_k} A$ in \eqref{projection} and integrate it with respect to time. 
We have:
\begin{eqnarray*}
&&\hspace{-1cm}
\left\langle A^{N_k}, B^{N_k}(t) \right\rangle 
- \left\langle A^{N_k}, B^{N_k}(0) \right\rangle 
+ \int_0^t \left\langle \nabla \times A^{N_k},(\nabla \times B^{N_k} )\times B^{N_k} \right\rangle \, ds +\\
&&\hspace{7cm}
+ \int_0^t \left\langle \nabla \times A^{N_k},\nabla \times B^{N_k} \right\rangle \, ds =0 . 
\end{eqnarray*}
Thanks to the Sobolev imbedding in dimension $3$, \, $\nabla \times A \in L^\infty(\Omega)^3$ \, and \,  $\nabla \times A^{N_k} \to \nabla \times A$ \, strongly in $L^\infty(\Omega)^3$. Now, thanks to the convergences (\ref{cvfaible}), we can take the limit $N_k \to \infty$ and get that $B$ satisfies
\begin{eqnarray*}
&&\hspace{-1cm}
\left\langle A, B(t) \right\rangle 
- \left\langle A, B(0) \right\rangle 
+ \int_0^t \left\langle \nabla \times A,(\nabla \times B )\times B \right\rangle \, ds +\\
&&\hspace{7cm}
+ \int_0^t \left\langle \nabla \times A,\nabla \times B \right\rangle \, ds =0, 
\end{eqnarray*}
which is a weak solution of the Hall problem. \endproof

\medskip
\noindent
{\bf Proof of theorem \ref{th:global}.} We apply the same Galerkin construction $(u^N, B^N)$ for the coupled system (\ref{eq:dtu}), (\ref{eq:dtB}) as we did for the Hall system and use the energy identity for the Galerkin approximation: 
\begin{eqnarray*}
& & \hspace{-1cm}
 \Vert u^N(t) \Vert ^2_{L^2(\Omega )} + \Vert B^N(t) \Vert ^2_{L^2(\Omega )} +2\int_0^t (  \Vert \nabla u^N(s) \Vert ^2_{(L^2(\Omega ))^9} +  \Vert \nabla B^N(s)\Vert ^2_{(L^2(\Omega ))^9} ) \,  ds = \\
 & & \hspace{8cm}
 = \Vert u^N_0\Vert ^2_{L^2(\Omega )} + \Vert B^N_0\Vert ^2_{L^2(\Omega )}.
\end{eqnarray*} 
Then the same proof can be reproduced. We just note that the time regularity of $u$ can be improved because there is no Hall term in the velocity equation. Therefore, we find $u_t \in  L^{\frac{4}{3}}((0,T),H^{-1}(\Omega))$ while $B_t \in  L^{\frac{4}{3}}((0,T),H^{-2}(\Omega))$. \endproof

%%%%%%%%%%%%%%%%%%%%%%%%%%%%%%%%%%%%%%%%%%%%%%%%%%%%%%%%%%%%%%%%%%%%%%%%%%%%%%%%%%%%%%%%%%%%%%%%
%%%%%%%%%%%%%%%%%%%%%%%%%%%%%%%%%%%%%%%%%%%%%%%%%%%%%%%%%%%%%%%%%%%%%%%%%%%%%%%%%%%%%%%%%%%%%%%%
%%%%%%%%%%%%%%%%%%%%%%%%%%%%%%%%%%%%%%%%%%%%%%%%%%%%%%%%%%%%%%%%%%%%%%%%%%%%%%%%%%%%%%%%%%%%%%%%
%%%%%%%%%%%%%%%%%%%%%%%%%%%%%%%%%%%%%%%%%%%%%%%%%%%%%%%%%%%%%%%%%%%%%%%%%%%%%%%%%%%%%%%%%%%%%%%%
\setcounter{equation}{0}
\section{Axisymmetric flows, KMC waves and Maxwell regularization of the non-resistive Hall problem}
\label{sec:axi}

In this section, we assume axisymmetric $B$ field. Let ${\mathbf x}= (x,y,z)$ a coordinate system, where $x$ is the symmetry axis. Axisymmetry about the $x$-axis means that, given any rotation $R$ about this axis, the field $B$ satisfies: $B(R{\mathbf x}) = R B({\mathbf x})$. We use $(x,r,\theta)$ the cylindrical coordinates of ${\mathbf x}$ and $(e_x, e_r, e_\theta)$ as associated local basis. 
Then, using a representation of axisymmetric, divergence-free fields given in \cite{Liu_Wang}, we can write 
$$ B = b e_\theta + \nabla \times (\psi e_\theta), $$
where the scalar functions $b$ and $\psi$ are functions of $(x,r)$. We note the simple formulas \cite{Liu_Wang}: 
\begin{eqnarray*}
& & \hspace{-1cm} 
\nabla \times B = - {\mathcal L} \psi e_\theta + \nabla \times (b e_\theta), \quad {\mathcal L} = \nabla_{(x,r)}^2 - \frac{1}{r^2}, \\
& & \hspace{-1cm} 
\nabla \times (\nabla \times B) = - {\mathcal L} b + \nabla \times (j e_\theta), \quad j= - {\mathcal L} \psi, \\
& & \hspace{-1cm} 
\nabla \times (( \nabla \times B) \times B) = \left( \left\{ \frac{j}{r}, r \psi \right\}_{x,r} - 
\left\{ \frac{b}{r}, r b \right\}_{x,r} \right) e_\theta + \\
& & \hspace{5cm} 
+ \nabla \times \left( \left( \frac{1}{r^2}
\left\{ rb , r \psi \right\}_{x,r} \right) e_\theta \right), 
\end{eqnarray*}
where the Poisson Bracket $\left\{ a, b \right\}_{x,r} = \partial_x a \, \partial_r b -  \partial_x b \,  \partial_r a$. 

In these coordinates, the Hall problem (\ref{eq:dtB2}), (\ref{eq:divB2}) is written:
\begin{eqnarray*}
& & \hspace{-1cm} 
\psi_t + \frac{1}{r^2} \left\{ rb , r \psi \right\} = {\mathcal L} \psi, \\
& & \hspace{-1cm} 
b_t + \left\{ \frac{j}{r}, r \psi \right\} - 
\left\{ \frac{b}{r}, r b \right\} = {\mathcal L} b. 
\end{eqnarray*}
If initially $\psi_0 = 0$ (which means that the $B$ field is purely swirling), then $\psi \equiv 0$ at all times and the $b$ equation reduces to the simple viscous Burger's equation
$$ b_t - \frac{2}{r} b \, b_x =  {\mathcal L} b. $$
In the inviscid case, the Burger's equation has solutions in the form of propagating shock waves. In the context of Hall MHD, these waves are known as the KMC waves, for Kingsep, Mokhov and Chukbar \cite{KMC} (see also \cite{CG}). They only exist if the Hall term is present.

Here, we focus on the non-resistive Hall problem and formulate it as a limit of a so-called Maxwell regularization. The non-resistive Hall problem is written
\begin{eqnarray*}
& & \hspace{-1cm} 
\partial_t B + \nabla \times E = 0, \\
& & \hspace{-1cm} 
\nabla \times B = j , \\
& & \hspace{-1cm} 
E = j \times B.  
\end{eqnarray*}
In the case of axisymmetric purely swirling $B$-fields, this yields the inviscid Burger's equation as seen 
above. We now consider a regularization of this problem by restoring the displacement current in the Ampere equation. This yields the problem
\begin{eqnarray}
& & \hspace{-1cm} 
\partial_t B + \nabla \times E = 0, \label{eq:MRfaraday}\\
& & \hspace{-1cm} 
- \varepsilon \partial_t E + \nabla \times B = j , \label{eq:MRampere}\\
& & \hspace{-1cm} 
E = j \times B,  \label{eq:Hall}
\end{eqnarray}
where $\varepsilon \ll 1$ is the Maxwell regularization parameter. We now investigate two formulations of the Maxwell-regularized problem. 

\medskip
\noindent
{\em 1. The $(B,j)$ formulation. } This formulation consists in classically eliminating $E$ for $j$ and gives 
\begin{eqnarray*}
& & \hspace{-1cm} 
\partial_t B + \nabla \times (j \times B) = 0, \\
& & \hspace{-1cm} 
- \varepsilon \partial_t (j \times B) + \nabla \times B = j .
\end{eqnarray*}
When $\varepsilon \to 0$, it clearly tends to the non-resistive Hall problem.

\medskip
\noindent
{\em 2. The $(B,E)$ formulation. } This formulation consists in eliminating $j$ for $E$. 
From (\ref{eq:Hall}), we can write
$$ j = \frac{B \times E}{|B|^2} + \lambda B, $$
where $\lambda$ is a priori unknown. However, we also have the geometric constraint
\begin{equation} 
E \cdot B = 0, 
\label{eq:EB}
\end{equation}
and $\lambda$ can be viewed as a Lagrange multiplier of this constraint. Indeed, from 
\begin{eqnarray*}
& & \hspace{-1cm} 
- \varepsilon \partial_t E + \nabla \times B = \frac{B \times E}{|B|^2} + \lambda B .
\end{eqnarray*}
and taking the scalar product of this equation with $B$, we get
\begin{eqnarray*}
& & \hspace{-1cm} 
- \varepsilon (\partial_t E) \cdot B + (\nabla \times B) \cdot B = \lambda |B|^2 .
\end{eqnarray*}
But, differentiating the constraint (\ref{eq:EB}) gives
\begin{eqnarray*}
& & \hspace{-1cm} 
(\partial_t E) \cdot B = - (\partial_t B) \cdot E = (\nabla \times E) \cdot E.
\end{eqnarray*}
Therefore, 
\begin{eqnarray*}
& & \hspace{-1cm} 
\lambda = \frac{1}{|B|^2} ( - \varepsilon (\nabla \times E) \cdot E + (\nabla \times B) \cdot B ) .
\end{eqnarray*}
Finally, the Maxwell-regularized Hall problem in the $(E,B)$ formulation reads 
\begin{eqnarray*}
& & \hspace{-1cm} 
\partial_t B + \nabla \times E = 0, \\
& & \hspace{-1cm} 
- \varepsilon \partial_t E + \nabla \times B = \frac{1}{|B|^2} (B \times E + (- \varepsilon (\nabla \times E) \cdot E + (\nabla \times B) \cdot B) \, B ) .
\end{eqnarray*}
In this system, the constraint (\ref{eq:EB}) is satisfied as soon as it is satisfied at $t=0$. 
The fact that the limit $\varepsilon \to 0$ of this problem leads to the Hall problem is no more so obvious. Indeed, the limit $\varepsilon \to 0$ leads to the following problem:
\begin{eqnarray*}
& & \hspace{-1cm} 
\partial_t B + \nabla \times E = 0, \\
& & \hspace{-1cm} 
\nabla \times B = \frac{1}{|B|^2} (B \times E +  ( (\nabla \times B) \cdot B ) \, B) , \\
& & \hspace{-1cm} 
\partial_t (E \cdot B ) = 0.
\end{eqnarray*}
The second equation is equivalent to 
$$ B \times \left[ (\nabla \times B) \times B - E \right] = 0. $$
If $(E\cdot B)|_{t=0} = 0$, then $E \cdot B \equiv 0$ for all times; Then, we can invert this equation into 
$$ E = (\nabla \times B ) \times B, $$ 
and recover the Hall problem as the formal limit of the Maxwell-regularized system. It is an interesting problem to determine if this limit can be made rigorous.

%%%%%%%%%%%%%%%%%%%%%%%%%%%%%%%%%%%%%%%%%%%%%%%%%%%%%%%%%%%%%%%%%%%%%%%%%%%%%%%%%%%%%%%%%%%%%%%%
%%%%%%%%%%%%%%%%%%%%%%%%%%%%%%%%%%%%%%%%%%%%%%%%%%%%%%%%%%%%%%%%%%%%%%%%%%%%%%%%%%%%%%%%%%%%%%%%
%%%%%%%%%%%%%%%%%%%%%%%%%%%%%%%%%%%%%%%%%%%%%%%%%%%%%%%%%%%%%%%%%%%%%%%%%%%%%%%%%%%%%%%%%%%%%%%%
%%%%%%%%%%%%%%%%%%%%%%%%%%%%%%%%%%%%%%%%%%%%%%%%%%%%%%%%%%%%%%%%%%%%%%%%%%%%%%%%%%%%%%%%%%%%%%%%
\setcounter{equation}{0}
\section{Conclusion}
\label{sec:conclu}

In this paper, we have derived and analyzed the Hall-MHD model. First, the model has been derived from a scaling limit of a two-fluids Euler-Maxwell system, under suitable scaling assumptions. Then, a derivation of the Hall-MHD from a kinetic model consisting of a Fokker-Planck equation for the ions coupled with fluid electrons has been realized. In the analysis section, we have proved the existence of global weak solutions for the incompressible viscous resistive Hall-MHD problem. The proof relies strongly on the skew-symmetric structure of the Hall term, which does not affect the energy estimate. This work shows that maintaining this structure is crucial for the well-posedness of the problem and is likely to be crucial as well for the derivation of stable numerical approximations. The question of the perfectly conducting boundary condition will be also investigated in future work. In the last section, we have reviewed some aspects of axisymmetric, purely swirling magnetic fields and proposed a regularization of the Hall problem by reintroducing the displacement current in Ampere's equation.  Future work will be devoted to the analysis of the axisymmetric case and to the Maxwell regularization of the Hall problem, as well as to the investigation of the kinetic formulations of the Hall-MHD problem.

%%%%%%%%%%%%%%%%%%%%%%%%%%%%%%%%%%%%%%%%%%%%%%%%%%%%%%%%%%%%%%%%%%%%%%%%%%%%%%%%%%%%%%%%%%%%%%%%
%%%%%%%%%%%%%%%%%%%%%%%%%%%%%%%%%%%%%%%%%%%%%%%%%%%%%%%%%%%%%%%%%%%%%%%%%%%%%%%%%%%%%%%%%%%%%%%%
%%%%%%%%%%%%%%%%%%%%%%%%%%%%%%%%%%%%%%%%%%%%%%%%%%%%%%%%%%%%%%%%%%%%%%%%%%%%%%%%%%%%%%%%%%%%%%%%
%%%%%%%%%%%%%%%%%%%%%%%%%%%%%%%%%%%%%%%%%%%%%%%%%%%%%%%%%%%%%%%%%%%%%%%%%%%%%%%%%%%%%%%%%%%%%%%%

\end{document}